\documentclass[preprint,preprintnumbers,amsmath,amssymb]{revtex4}

\usepackage{graphicx}% Include figure files
\usepackage{dcolumn}% Align table columns on decimal point
\usepackage{bm}% bold math

\begin{document}

\newcommand*{\cm}{cm$^{-1}$\,}

%\preprint{APS/123-QED}

\title{Structural phase transition in IrTe$_2$: A combined study of optical spectroscopy and band structure calculations}

\author{A. F. Fang}
\author{G. Xu}
\author{T. Dong}
\author{P. Zheng}
\author{N. L. Wang\footnote{Correspondence and requests
for materials should be addressed to N.L.W.
(nlwang@aphy.iphy.ac.cn).}}
\affiliation{Beijing National Laboratory for Condensed Matter
Physics, Institute of Physics, Chinese Academy of Sciences,
Beijing 100190, People's Republic of China}

%


\begin{abstract}

Ir$_{1-x}$Pt$_x$Te$_2$ is an interesting system showing competing phenomenon between
structural instability and superconductivity. Due to the large
atomic numbers of Ir and Te, the spin-orbital coupling is expected to be strong in the system
which may lead to nonconventional superconductivity.
We grew single crystal samples of this
system and investigated their electronic properties. In particular, we performed optical spectroscopic
measurements, in combination with density function calculations, on the undoped compound IrTe$_2$ in an effort to elucidate the origin of the structural phase transition at 280 K. The measurement revealed a dramatic reconstruction of band structure and a significant reduction of conducting carriers below the phase transition. We elaborate that the transition is not driven by the density wave type instability but caused by the crystal field effect which further splits/separates the energy levels of Te (p$_x$, p$_y$) and Te p$_z$ bands.

\end{abstract}


\maketitle

Transition metal dichalcogenides are layered compounds exhibiting a
variety of different ground states and physical properties. Among
the different polytypes, the compounds with 1T and 2H structural
forms TX$_2$ (T=transition metal, X=chalcogen) have attracted
particular attention. Those compounds consist of stacked layers of
2D-triangular lattices of transition metals (e.g., Ti, Ta or Nb)
sandwiched between layers of chalcogen atoms (e.g., S, Se or Te) with
octahedral and trigonal prismatic coordinations,
respectively \cite{1T2Hstructure,wilson1975,TaSe2NbSe2Neutron}. The low dimensionality of those
systems often leads to charge density wave (CDW) instability \cite{wilson1975,TaSe2NbSe2Neutron,TX2CDWSC,Valla,Borisenko,Morosan,TiSe2,Zhao,WZHu}. Yet, a number of those CDW-bearing materials are also superconducting \cite{TX2CDWSC,Valla,Borisenko,Morosan}. The
interplay between the two very different cooperative electronic
phenomena is one of the fundamental interests of condensed-matter
physics. Recently, the 5d transitional metal ditelleride IrTe$_2$ has added to the interest. The compound has a layered 1T structure consisting of edge-sharing IrTe$_6$ octahedra (see Fig. 1 (a)-(c)), and exhibits a structural phase transition near 270 K \cite{resistance}. The transition leads to sharp changes in resistivity and magnetic susceptibility. It was found very recently that both Pt or Pd intercalations and substitutions could induce bulk superconductivity with T$_c$ up to $\sim$ 3 K \cite{polysample,JJYang,XPS,thermalconductivity}. Since both Ir and Te possess large atomic numbers, strong spin-orbital coupling must be present in those compounds, which may induce topologically nontrivial state. There is a possibility that the Pt or Pd intercalated or substituted compounds are better candidates for topological superconductors due to its improved superconductivity in comparison with Cu$_x$Bi$_2$Se$_3$\cite{CuBiSe1,CuBiSe2,CuBiSe3}.

Understanding the structural phase transition is a crucial step towards understanding the electronic properties of the
system. Up to now, there have been very limited studies on the origin of the structural phase transition near 270 K. Structural characterizations indicated that the transition is accompanied by the appearance of superstructure peaks with wave vector of \textbf{q}=(1/5, 0, -1/5), which may match with the Fermi surface (FS) nesting wave vector from theoretical calculations \cite{JJYang}. Therefore, the transition was suggested to be a CDW phase transition with involvement of orbital degree \cite{JJYang}, and the superconductivity competed with the CDW in a quantum critial point-like manner. On the other hand, a recent Ir 4f XPS study on IrTe$_2$ suggested that the low temperature phase of IrTe$_2$ was accompanied by the establishment of weak modulation of Ir
5d t$_{2g}$ electron density and the structural phase transition of IrTe$_2$ was argued to be an orbitally-induced Peierls transition \cite{XPS}, being similar to the structural transition of spinel-type CuIr$_2$S$_4$ \cite{Khomskii}.

In this work, we present a detailed optical spectroscopic study on well characterized Ir$_{1-x}$Pt$_{x}$Te$_2$ ($x$= 0, 0.05) single crystal samples. We find that the structural phase transition in the undoped compound is associated with the sudden reconstruction of band structure over a broad energy scale up to at least 2 eV. Although the carrier density is significantly reduced after the phase transition, yielding evidence for a reduced FS area, we could not identify any energy gap associated with the transition with a characteristic lineshape from the coherent factor of density wave collective phenomenon. In combination with the density function calculations, we elaborate that the structural phase transition is not of a density wave type but of a different origin. A novel explanation is proposed to explain the structural instability of this interesting material in terms of experimental and band structural calculation results.

\begin{flushleft}
    \textbf{Results}
\end{flushleft}

Figure 1 (a)-(c) show the structural characteristics of IrTe$_2$ compound.
Pictures of plate-like IrTe$_2$ single crystals with
maximum $ab$ plane dimension of 6 $\times$ 4 mm$^{2}$ are shown in the
inset of the figure. Fig. 1 (d) displays the X-ray diffraction patterns of IrTe$_2$ single
crystals at room temperature. The (0 0 $l$)
diffraction peaks indicate a good \emph{c}-axis characteristic. Several
single crystals were pulverized to measure powder XRD [Fig. 1 (e)].
The major peak patterns confirm a single trigonal phase with the
lattice parameters $a$=3.9322(5) ${\AA}$ , $c$=5.3970(8) ${\AA}$ with
space group $P\bar 3 m1$, which is consistent with the previous
report \cite{bandcaculation,resistance}. Compared with pure IrTe$_2$, the major peaks of Ir$_{0.95}$Pt$_{0.05}$Te$_2$ shift toward lower 2$\theta$, while they could be indexed by the same structure, resulting in slightly different lattice parameters: longer $a$=3.9394(1) ${\AA}$ and shorter $c$=5.3890(5) ${\AA}$. The evolution of the lattice parameters is quite consistent with the previous report on polycrystalline samples \cite{polysample}. The minor extra peak denoted by asterisks
were determined to be Te phase, possibly from some Te flux remaining
on the crystal surfaces.

Figure 2 presents the physical properties characterizations of Ir$_{1-x}$Pt$_{x}$Te$_2$ single crystals.
Figure 2 (a) displays the normalized dc resistivity $\rho/\rho_{300K}$ for Ir$_{1-x}$Pt$_{x}$Te$_2$ over broad temperature range (2 K $\sim$ 400 K). The most striking phenomenon for pure IrTe$_2$  is the steep jump
at 271 K (reaching maximum at 261 K) on cooling, 285 K (maximum at 275 K)
on heating. The significant hysteresis suggests a first order phase transition. The resistivity $\rho$(T) keeps
decreasing with lowering temperature after the phase transition. An
upward curvature is seen below 50 K, reminiscent of ordinary
Bloch-Gruneisen lineshape due to electron-phonon interactions. However, after 3\% Pt doped into IrTe$_2$, the transition temperature suddenly descends to 130 K (cooling process); With 5\% Pt doping, the phase transition is completely absent. Above the transition temperatures all the $\rho/\rho_{300K}$ resistivity curves could be perfectly scaled together. We
find that, while some of the pure IrTe$_2$ crystals just show flat residual
resistivity at low temperature as shown in Fig. 2 (a), some others
grown from the same batch exhibit sharp drops at 2.5 K in
$\rho$(T) curve. We applied magnetic field to the latter and found
that the transition temperature reduced to 1.5 K under 1 Tesla with
field parallel to $ab-$plane, as shown in Fig. 2 (d). The feature is
analogous to the resistivity behavior of superconductivity.
Nevertheless, no anomaly could be identified in the specific heat
measurement down to 0.5 K [the inset of Fig.2 (c)]. The measurement
revealed filament superconductivity in some of the crystals,
possibly originated from Ir vacancies or excess Te in the samples.
Bulk superconductivity is seen in Pt doped samples, similar to polycrystalline
samples \cite{polysample}. Figure 2(e) shows the low temperature magnetic susceptibility
data of Ir$_{0.95}$Pt$_{0.05}$Te$_2$, indicating strong superconducting diamagnetic effect. Simultaneously the minor distinction between FC and ZFC processes implies weak superconducting vortex pinning, which, to a certain extent, indicates the good quality of our single crystals. A significant jump in the specific heat measurement at $T_c$ is observed for Ir$_{0.95}$Pt$_{0.05}$Te$_2$, as shown in Fig. 2(f).

The magnetic susceptibilities of Ir$_{1-x}$Pt$_{x}$Te$_2$ single crystals in an applied magnetic field H = 1 T aligned in $ab-$plane are illustrated in Fig. 2(b). For samples with Pt contents $x$= 0 and 0.03, the temperature-independent diamagnetism is observed above their separate structural phase transitions, and then sudden drops happen at the transition temperatures, resulting in more diamagnetic behavior. For Ir$_{0.95}$Pt$_{0.05}$Te$_2$, the magnetic susceptibility practically keeps constant except for the upturn at low temperature which is likely from paramagnetic impurities. The observation of diamagnetic susceptibilities even above the structural phase transition is different from the several recent reports on polycrystalline samples \cite{resistance,JJYang,polysample}. In an earlier report, diamagnetism was observed in the entirely measured temperature range below 300 K, however no phase transition could be seen in the experimental data \cite{bandcaculation}. We emphasize that our measurements are performed on single crystal samples and are more reliable. As we shall see below from optical measurement, all samples are metallic both above and below structural phase transitions, the observation of diamagnetism is rather surprising. This is because, for metallic compounds, the diamagnetism is usually a weak effect and the Pauli paramagnetism should dominate the magnetic susceptibility. As we shall explain below, this peculiar property is related to the valence state of Ir$^{3+}$ which leads to more closed shells.

We performed the specific heat measurements over broad temperature range (0.5 K $\sim$ 300 K) for the compounds. As illustrated in Fig. 2 (c), a very pronounced sharp peak appears near 280 K for the undoped compound. The
characteristic $\delta$-like shape of the peak reveals a first
order phase transition. In the low temperature range, the specific
heat follows the relation of $C_p = \gamma T + \beta {T^3}$. From a
plot of $C/T$ versus $T^{2}$, shown in the inset of Fig. 2(c), we get
$\gamma$=3.96(3) mJK$^{-2}$mol$^{-1}$  and $\Theta _ D$=151 K. Those values are for the
low temperature phase. For the Pt-doped superconducting sample with $x=0.05$, we obtain $\gamma$=7.02(7) mJK$^{-2}$mol$^{-1}$  and $\Theta _ D$=147 K based on specific heat data just above the superconducting
transition temperature [Fig. 2(f)]. It is known that $\gamma$ is proportional to the electronic density of states near the fermi level. Taking $\gamma$ = 3.96 mJK$^{-2}$mol$^{-1}$ in the low-T distorted structural phase of pure IrTe$_2$ and assuming that the $\gamma$ value in the high-T trigonal phase is approximated to be the same as that of Ir$_{0.95}$Pt$_{0.05}$Te$_2$
(i.e. ignore the change of the electronic specific heat coefficient due to Pt doping),
we estimate that roughly 44\% fermi surface is removed across the structural phase transition.

Below we shall focus on the structural phase transition near 280 K.
Fig. 3 (a) presents the reflectance curves below 27500 \cm. The
inset is the expanded low frequency range below 3000 \cm. All the
spectra approach to unity at zero frequency and good metallic
response can be seen both above and below the phase transition.
In both high and low temperature phases, R($\omega$) shows relatively small
change with a variation of temperature.
However, a huge spectral change occurs on cooling the sample across the first
order phase transition at 280 K. The reflectance values below roughly 15000 \cm in the high
temperature phase is significantly higher than those in the low temperature phase, suggesting
much higher Drude spectral weight in the high temperature phase. This is seen
clearly from the low frequency optical conductivity spectra shown in Fig. 3 (b).

Since the spectral weight of the Drude component is proportional to the square of the
plasma frequency $ \omega _p^2 = 4\pi ne^2/m^*$, the sudden reduction of the Drude spectral
weight reflects a reduction of the $n/m^*$. Besides the Drude component, the interband transitions at higher energies seem to experience some change as well. For example, the clear interband transition peak near 15000 \cm at high temperature phase becomes obscured below the phase transition. The overall
spectral change reflects a significant band structure reconstruction associated with the first order phase transition.

To quantify the spectral change, particularly the evolution of the Drude
component, across the phase transition, we tried to decompose the optical conductivity
spectral into different components using a Drude-Lorentz analysis.
The dielectric function has the form \cite{TiSe2}
\begin{equation}
\epsilon(\omega)=\epsilon_\infty-\sum_{i}{{\omega_{p,i}^2}\over{\omega_i^2+i\omega/\tau_i}}+\sum_{j}{{\Omega_j^2}\over{\omega_j^2-\omega^2-i\omega/\tau_j}}.
\label{chik}
\end{equation}
where $\epsilon_\infty$ is the dielectric constant at
high energy,  and the middle and last terms are the Drude and Lorentz components,
respectively. The Drude component represents the contribution from conduction
electrons, while the Lorentz components describe the interband
transitions. We found that the optical conductivity spectra below
25000 \cm could be reasonably reproduced by one Drude and four
Lorentz components. Fig. 3 (c) shows the conductivity spectra at 300
K and 250 K together with the Drude-Lorentz fitting components. For
clarity, the Lorentz components only at 300 K were shown in the
figure. The parameters of Drude components above and below the
structural phase transition, \emph{e.g.} at 300 K and 250 K, are
$\omega _P\approx$ 39000 cm$^{-1}$, 29200 cm$^{-1}$ and $1/\tau\approx$
900 cm$^{-1}$, 1250 cm$^{-1}$, respectively. The inset of Fig.3 (c)
displays the temperature-dependent evolution of $1/\tau_j$ and $\omega _P^2$. Both
parameters are normalized to the values of 350 K. The sudden decease
of plasma frequency is seen only at the phase transition. At other temperatures
the plasma frequencies keep roughly unchanged, but the
scattering rates decrease with decreasing temperature in both phases, reflecting
the expected narrowing of Drude component for a metallic response.
The scattering rate at 250 K is higher than the value at 300 K, which is
consistent with the observation of dc resistivity measurement
showing a higher resistivity near 250 K after the jump at the
transition. It is noted that the ratio of the square of the
plasma frequency at low temperature phase to that at high
temperature phase is about 0.56. Provided that the effective mass of
the itinerant carriers remains unchanged, the measurement reveals
that roughly 44\% itinerant carriers are lost after the first order
phase transition. This is in agreement with our specific heat estimation and the NMR measurement result \cite{NMR}.
The suppressed Drude spectral weight is
transferred to higher energies due to the band reconstruction. The
reduced spectral weight is recovered roughly at the frequency of
25000 \cm.

Fig. 4 displays the reflectance and conductivity spectra for Ir$_{0.95}$Pt$_{0.05}$Te$_2$ single crystals with the variation of temperature from 10 K to 300 K. Because the structural transition is completely suppressed, the huge spectral change seen for the pure IrTe$_2$ across the structural phase transition is apparently absent for Ir$_{0.95}$Pt$_{0.05}$Te$_2$. The low frequency Drude component shows usual narrowing due to the reduced scatterings with decreasing temperature. Weak but visible intensity change could be identified for the interband transition peaks at high frequencies. Those interband transitions may involve the bands across the Fermi level. Due to the effect of temperature-dependent Fermi distribution function, the electron occupations near the Fermi level on those bands would show a small change at different temperatures. This may explain the observed intensity change.

\begin{flushleft}
    \textbf{Discussion}
\end{flushleft}

Understanding the origin of the structural phase transition is a key to the understanding
of electronic properties of the system. The transition
was suggested to be a kind of charge density wave with some involvement of
Ir 5d orbitals \cite{JJYang}. However, our present measurement suggests against a density wave
type phase transition. It is known that a hallmark of
the density-wave type transition is the formation of an energy gap in the
single particle excitation spectrum near the Fermi level \cite{densitywaves}, resulting
in the lowering of the total energy of the system. Within the scheme of the
BCS theory for density wave condensate, the opening of energy gap
leads to a characteristic peak feature
just above the energy gap in optical conductivity due to the effect of so-called
"type-I coherent factor" \cite{densitywaves}. The energy scale of the gap is
related to the transition temperature, for example,
2$\Delta/T_{DW}\sim$ 3.5 under the weak-coupling BCS theory. However, in the present case,
no such energy gap at low energy, or in the energy range with much larger gap
value, could be identified from the optical conductivity spectra.
As we discussed above, the spectral change
occurs over a very broad energy range up to 25000 \cm, which is
attributed to the reconstruction of the band structure associated
with the phase transition. Furthermore, according to all available structural
investigation, all Ir sites are still equivalent even in the low temperature low symmetry
phase \cite{resistance}. As a result, the structural phase transition is
not likely to be driven by the so-called orbital-driven Peierls transition as well. A
recent angle-resolved photoemission spectroscopy (ARPES) study is in
agreement with the present conclusion. \cite{Ootsuki}

To gain insight into the nature of the structural phase transition, we have carefully
examined the structural characteristic and its change across the phase transition and performed
local density approximation (LDA) band structure calculations. A very remarkable and
characteristic structural feature about IrTe$_2$ compound is that
the average Te-Te bond length d$_{Te-Te}$ = 3.528 ${\AA}$ is noticeably shorter
than the usual value of 4.03 ${\AA}$ of regular Te$^{2+}$ in classic CdI$_2$-like
arrangement (e.g. HfTe$_2$), making the valence state of Ir close to Ir$^{3+}$
rather than Ir$^{4+}$ \cite{bandcaculation}. The charge balance of this compound
is close to Ir$^{3+}$[Te$^{-1.5}]_2$, \emph{i.e.} the Iridium
ion is close to 5d$^6$. Another piece of important information yielded from the
available structural investigation is that the
structural distortion below the phase transition mainly causes a reduction of
the Te-Te bond length between the upper and lower
planes of the IrTe$_6$ octahedra slabs from 3.528 ${\AA}$ to 3.083 ${\AA}$
(i.e. d$_3$ in Fig. 1 (c)).
The change of Te-Te bond length within the Te layers of each IrTe$_6$ octahedra slab
across the transition is much smaller, from 3.928 ${\AA}$
to 3.934 ${\AA}$ and 3.812 ${\AA}$ (i.e. d$_1$ and d$_2$ in Fig. 1 (c)), respectively  \cite{resistance}. Those structural
characteristics provide hint on the origin of the
structural phase transition.

In the following we shall propose a novel interpretation for the structural phase transition.
Since the Ir sits in the octahedral crystal field formed by Te, the Ir 5d level would
split into t$_{2g}$ (d$_{xy}$, d$_{yz}$, d$_{zx}$) and e$_g$ (d$_{x^2-y^2}$, d$_{z^2}$) manifolds.
The six 5d$^6$ electrons would mainly occupy the t$_{2g}$ levels,
leading to almost fully filled t$_{2g}$ bands.
Although there must be some hybridizations between Ir 5d and Te 5p orbitals, the states near the
Fermi level should be dominantly contributed by the Te 5p orbitals. Indeed, this was confirmed by the local-density approximation (LDA) band structure calculations as presented in Fig. 5. Therefore, as a simplified picture, we shall only consider the Te 5p (p$_x$, p$_y$, p$_z$) orbitals which accommodate roughly 5.5 electrons. Figure 5 (a) and (c) show the band dispersions and density of state (DOS) of IrTe$_2$ using the experimentally determined crystal structures at high temperature phase \cite{resistance}. In the calculations, we have taken
into account of the spin-orbital coupling and applied the generalized gradient approximation (GGA) for the exchange-correlation potential. There are two bands crossing Fermi level if looking along $\Gamma$-M line. One is closer to $\Gamma$ which is contributed dominantly from the Te 5p$_x$ and 5p$_y$ orbitals and is largely filled; the other one, being contributed dominantly from Te 5p$_z$ orbtial, is more dispersive and crosses E$_F$ at higher momentum position. The resultant two FSs at high temperature phase are plotted in Fig. 5 (e). The results are in agreement with the recently reported work \cite{JJYang}. However, our calculated Lindhard response function does not show strong features at any wave vector. The FS itself is not likely to drive a density wave type instability.

Because the structural phase transition mainly causes a suppression of the IrTe$_6$ octahedral along the \emph{c}-axis, it is expected that the structural distortion would further split/separate the Te p$_z$ band from the Te (p$_x$, p$_y$) band (a schematic picture is illustrated in the inset of Fig. 5 (d)). The Te (p$_x$, p$_y$) band is further lowered and filled, while the Te p$_z$ band is largely broadened and less occupied. As a consequence, both the inner and outer FSs would become smaller. This is indeed the case in our LDA calculation with a shortened Te-Te bond length of d$_3$ in terms of the available crystal structure at low temperature phase \cite{resistance}, as shown in Fig. 5 (b), (d) and (f). Our calculations also indicate a reduction of the density of sate (DOS) at Fermi level from N(E$_F$)=2.161 to 1.335 states/eV f.u., \emph{i.e.} roughly a 40 $\%$ loss, which could qualitatively interpret the specific heat and optical results. The increased occupation of the further lowered Te (p$_x$, p$_y$) energy levels would result in a decrease of the kinetic energy of the electrons, which should be the driving force for the transition.

Finally, we comment on the striking diamagnetism observed in magnetic susceptibility measurement for those samples. As mentioned above, the valence state of Ir is Ir$^{3+}$, the 5d$^6$ electrons of Ir would fully fill the t$_{2g}$ bands which are split from the  e$_g$ bands in the octahedral crystal field. The conducting electrons are mainly from the Te 5p orbitals. It is well-known that the closed shells (or fully occupied bands) would contribute to the diamagnetism (Larmor diamagnetism). Because both Ir and Te are relatively heavy, many closed shells are present for the compound. As a result, the Pauli paramagnetism contributed from the Te 5p orbitals could not overcome the diamagnetism resulted from the closed shells. Below the structural phase transition, the Pauli paramagnetism is reduced due to a reduction of density of state near E$_F$, which results in further enhanced diamagnetism.

To conclude, we have successfully grown single crystal samples of Ir$_{1-x}$Pt$_{x}$Te$_2$ ($x$= 0, 0.03 and 0.05) and characterized their electronic properties. In particular, we performed a combined optical spectroscopy and first principle calculation study on the undoped sample of IrTe$_2$ in an effort to understand the origin of the
structural phase transition at 280 K. The measurement revealed a sudden reconstruction of band structure over broad energy scale and a significant removal of conducting carriers below the transition. The study indicated that the first order structural phase transition was not driven by the density wave type Fermi surface instability, but caused by the crystal field effect which further split/separated the energy levels of Te (p$_x$, p$_y$) and Te p$_z$ and resulted in a reduction of the kinetic energy of the electronic system.

\begin{flushleft}
    \textbf{Methods}
\end{flushleft}
\small{Single crystals of Ir$_{1-x}$Pt$_x$Te$_2$ have been successfully grown via
self-flux technique. The mixtures of Ir (Pt) powder and Te pieces in an
atomic ratio of 0.18:0.82 were placed in an Al$_2$O$_3$ crucible and
sealed in an evacuated quartz tube. The mixture was heated up
initially to 950 $^\circ \mathrm{C}$ and kept for several hours,
then to 1160 $^\circ \mathrm{C}$ for one day, and finally cooled
down slowly to 900 $^\circ \mathrm{C}$ at a rate of 2 $^\circ
\mathrm{C}$/$h$. The flux Te was separated from single crystals by
using a centrifuge. The dc resistivity data were measured with a commercial Quantum Design system PPMS by a four-probe method. The magnetic susceptibility was performed in a quantum design superconducting quantum interference device vibrating sample magnetometer system (SQUID-VSM). The specific-heat measurements were conducted by a relaxation-time method using PPMS. The temperature-dependent optical reflectance measurements were performed on Bruker 113v, Vertex 80v and a grating spectrometers on freshly cleaved surfaces of IrTe$_2$
single crystals. An
\textit{in situ} gold and aluminum over-coating technique was used
to get the reflectivity R($\omega$). The real part of conductivity $\sigma_1(\omega)$ is
obtained by the Kramers-Kronig transformation of R($\omega$).}

This work was supported by the National Science Foundation of
China, and the 973 project of the Ministry of Science and Technology of China (2011CB921701,2012CB821403).

\bibliographystyle{apsrev4-1}

\begin{references}

\bibitem{1T2Hstructure}  Wilson, J. A. and Yoffe, A. D. The transition metal dichalcogenides discussion and interpretation of the observed optical, electrical and structural properties. \emph{Adv. Phys.} \textbf{18}, 193 (1969).
\bibitem{wilson1975}  Wilson, J. A. \emph{et al.} Charge-density waves and superlattices in the metallic layered transition metal dichalcogenides.  \emph{Adv. Phys.} \textbf{24}, 117 (1975).
\bibitem{TaSe2NbSe2Neutron} Moncton, D. E. \emph{et al.} Neutron scattering study of the charge-density wave transitions in 2H-TaSe$_2$ and 2H-NbSe$_2$. \emph{Phys. Rev. B} \textbf{16}, 801 (1977).
\bibitem{TX2CDWSC} Castro Neto, A. H. Charge Density Wave, Superconductivity, and Anomalous Metallic Behavior in 2D Transition Metal Dichalcogenides. \emph{Phys. Rev. Lett.} \textbf{86}, 4382 (2001).
\bibitem{Valla} Valla, T. \emph{et al.} Quasiparticle Spectra, Charge-Density Waves, Superconductivity, and Electron-Phonon Coupling in 2H-NbSe$_2$. \emph{Phys. Rev. Lett.} \textbf{92}, 086401 (2004).


\bibitem{Borisenko} Borisenko, S. V. \emph{et al.} Two Energy Gaps and Fermi-Surface "Arcs" in NbSe$_2$.
 \emph{Phys. Rev. Lett.} \textbf{102}, 166402 (2009).



\bibitem{Morosan} Morosan, E. \emph{et al.} Superconductivity in Cu$_x$TiSe$_2$. \emph{Nature Phys}. \textbf{2}, 544 (2006) .
\bibitem{TiSe2} Li, G. \emph{et al.} Semimetal-to-Semimetal Charge Density Wave Transition in 1T-TiSe$_2$. \emph{Phys.Rev.Lett.} \textbf{99}, 027404 (2007).
\bibitem{Zhao} Zhao, J. F. \emph{et al.} Evolution of the Electronic Structure of 1T-Cu$_x$TiSe$_2$. \emph{Phys. Rev. Lett.} \textbf{99}, 146401 (2007).
\bibitem{WZHu} Hu, W. Z. \emph{et al.} Optical study of the charge-density-wave mechanism in 2H-TaS$_2$ and Na$_x$TaS$_2$. \emph{Phys. Rev. B} \textbf{76}, 045103 (2007).
\bibitem{resistance} Matsumoto, N. \emph{et al.} Resistance and Susceptibility Anomalies in IrTe$_2$ and CuIr$_2$Te$_4$. \emph{J. Low Temp. Phys.} \textbf{117}, 1129 (1999).
\bibitem{polysample} Pyon, S. \emph{et al.} Superconductivity Induced by Bond Breaking in the Triangular Lattice of IrTe$_2$. \emph{J. Phys. Soc. Jpn.} \textbf{81}, 053701 (2012) .
\bibitem{JJYang} Yang, J. J. \emph{et al.} Charge-Orbital Density Wave and Superconductivity in the Strong Spin-Orbit Coupled IrTe$_2$:Pd. \emph{Phys. Rev. Lett.} \textbf{108}, 116402 (2012).
\bibitem{XPS} Ootsuki, D. \emph{et al.} Orbital degeneracy and Peierls instability in the triangular-lattice superconductor Ir$_1-x$Pt$_x$Te$_2$. \emph{Phys. Rev. B} \textbf{86}, 014519 (2012).
\bibitem{thermalconductivity} Zhou, S. Y. \emph{et al.} Nodeless superconductivity in Ir$_{1-x}$Pt$_x$Te$_2$ with strong spin-orbital coupling. \emph{arXiv}:1209.4229.
\bibitem{CuBiSe1} Hor, Y. S. \emph{et al.} Superconductivity in Cu$_x$Bi$_2$Se$_3$ and its Implications for Pairing in the Undoped Topological Insulator.  \emph{Phys. Rev. Lett.} \textbf{104}, 057001 (2010).
\bibitem{CuBiSe2} Kriener, M. \emph{et al.} Bulk Superconducting Phase with a Full Energy Gap in the Doped Topological Insulator Cu$_x$Bi$_2$Se$_3$.  \emph{Phys.
    Rev. Lett.} \textbf{106}, 127004 (2011).
\bibitem{CuBiSe3} Fu, L. and Berg, E. Odd-Parity Topological Superconductors: Theory and Application to Cu$_x$Bi$_2$Se$_3$. \emph{Phys. Rev. Lett.} \textbf{105}, 097001 (2010).
\bibitem{Khomskii} Khomskii, D. I. and Mizokawa, T. Orbitally Induced Peierls State in Spinels. \emph{Phys. Rev. Lett.} \textbf{94}, 156402 (2005).


\bibitem{bandcaculation} Jobic, S. \emph{et al.} Crystal and electronic band structure of IrTe$_2$: Evidence of anionic bonds in a CdI$_2$-like arrangement. \emph{Z. anorg. allg. Chem.} \textbf{598}, 199 (1991).
\bibitem{NMR} Mizuno, K. \emph{et al.} 125Te NMR study of IrTe$_2$. \emph{Physica B: Condensed Matter} \textbf{312}, 818 (2002).

\bibitem{densitywaves} Gr\"{u}ner, G. \emph{Density Waves in Solids (Addison-Wesley, Reading, MA, 1994)}.


\bibitem{Ootsuki} Ootsuki, D. \emph{et al.} Electronic structure reconstruction by orbital symmetry breaking in IrTe$_2$. arXiv:1207.2613.


\end{references}

\begin{center}
\small{\textbf{Author contributions}}
\end{center}
\small{A.F.F. grew single crystals and carried out measurements. G.X. performed density function calculations.
T.D and P.Z. helped with transport and specific heat measurement. A.F.F. and N.L.W. wrote the
paper. N.L.W. supervised the project.}

\begin{center}
\small{\textbf{Additional Information}}
\end{center}
\small{The authors declare that they have no competing financial interests.}

\begin{figure}[b]
\includegraphics[clip,width=3.2in]{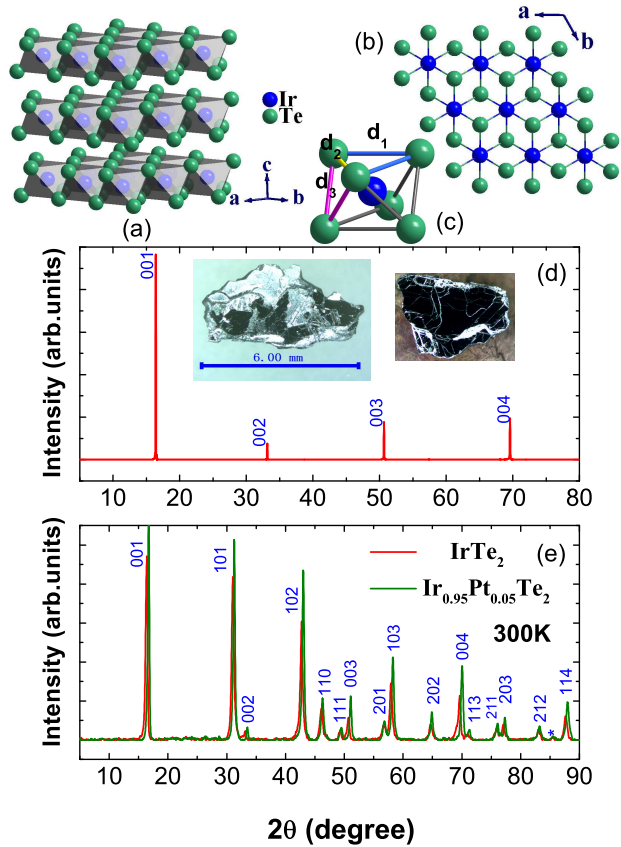}
\caption{(Color online) (a) Crystal structure of IrTe$_2$;
(b) The $ab$-plane structure of IrTe$_2$; (c) The IrTe$_6$ octahedral;
(d) X-ray diffraction patterns of IrTe$_2$
single crystals at room temperature . Inset: single crystal
pictures. (e) Powder XRD patterns of pulverized single crystals
and indexing. The peaks labeled by asterisks are from impurity
phase flux Te.}
\end{figure}

\begin{figure}
%\scalebox{0.35} {\includegraphics [bb=590 15 8cm 18cm]{Resistivity1.eps}}
\includegraphics[clip,width=5in]{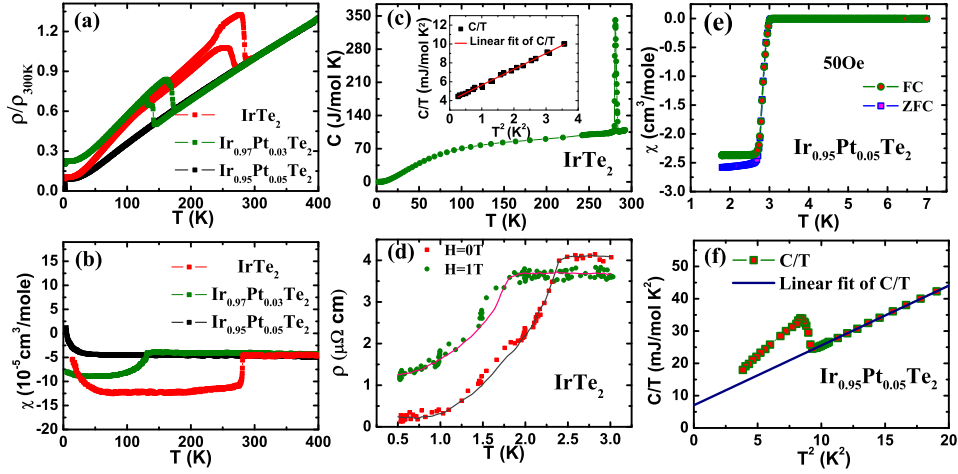}
\caption{(Color online) Physical properties characterizations of Ir$_{1-x}$Pt$_{x}$Te$_2$ single crystals. (a) The temperature dependence of normalized resistivity $\rho/\rho_{300K}$ for Ir$_{1-x}$Pt$_{x}$Te$_2$ ($x$= 0.0, 0.03 and 0.05). (b) The temperature dependent magnetic susceptibility of Ir$_{1-x}$Pt$_{x}$Te$_2$ with $x$= 0.0, 0.03 and 0.05, from 2 K to 400 K in a field of 1 T, being perpendicular to the $c$ axis. (c) Specific heat of IrTe$_2$ versus temperature. Inset shows the $C/T$ versus $T^{2}$ plot at low temperature. (d) The temperature dependence of resistivity $\rho$ from 0.5 K to 3 K with magnetic field parallel to \emph{ab}-plane for IrTe$_2$. The solid curves are guided to the eyes. (e) Low temperature magnetic susceptibility $\chi$ versus temperature for Ir$_{1-x}$Pt$_{x}$Te$_2$ at $x$=0.05 with applied magnetic field $H=50$ Oe parallel to \emph{ab}-plane. (f) Specific heat of Ir$_{0.95}$Pt$_{0.05}$Te$_2$ in the plot of $C/T$ versus $T^2$ with zero applied field; the solid line represents the linear fit of $C/T$ above the superconducting transition temperature $T_c$=3 K.}
\end{figure}

\begin{figure}
%\scalebox{0.68} {\includegraphics [bb=580 20 8cm 20cm]{optical1.eps}}
\includegraphics[clip,width=3in]{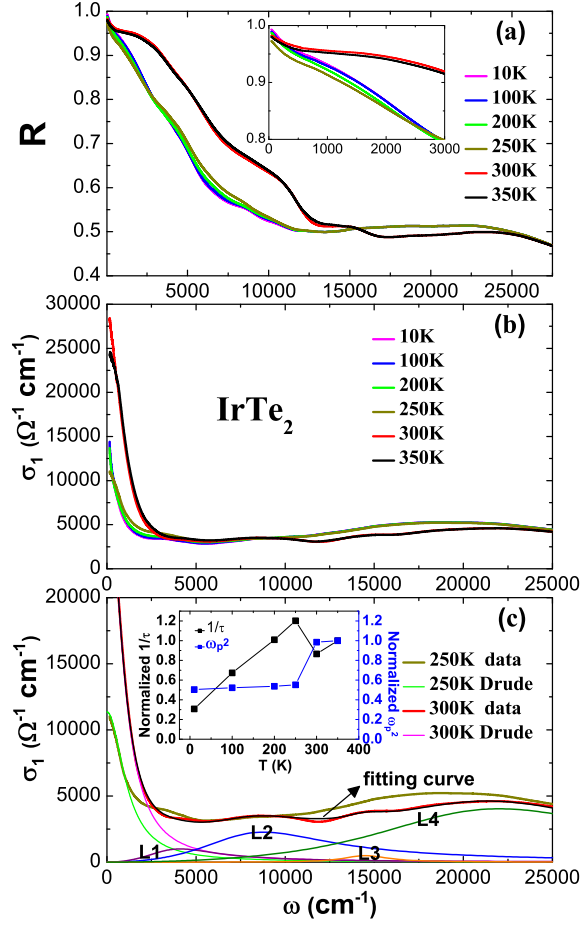}
\caption{(Color online) (a) The reflectance
 curves of IrTe$_2$ single crystals; Inset: R($\omega$) below
 3000 \cm. (b) The temperature dependence of the real part of the optical
conductivity $\sigma_1(\omega)$ for IrTe$_2$ single crystals up to
25000 \cm. (c) The experimental data of $\sigma_1(\omega)$ at 250K
and 300K with the Drude-Lorentz fits shown at the bottom.}
\end{figure}

\begin{figure}[t]
\includegraphics[clip,width=3in]{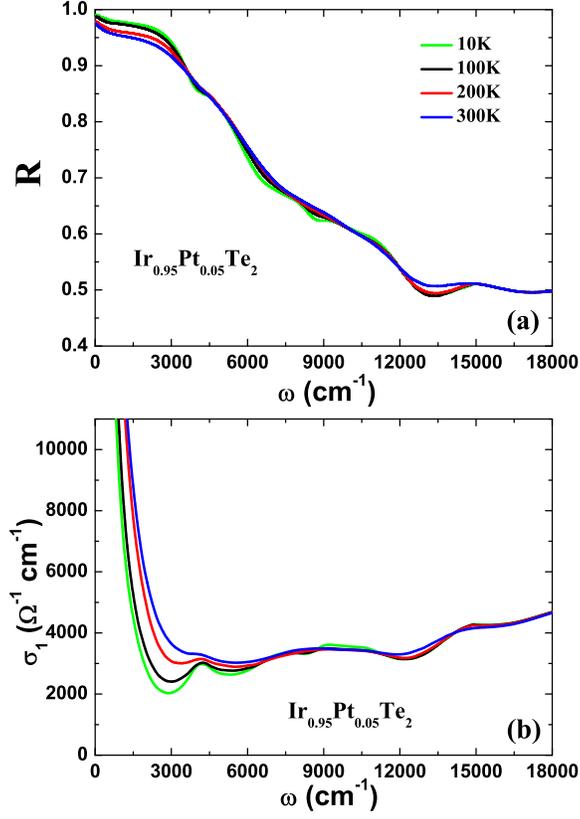}
\caption{(Color online) (a) The \emph{ab}-plane reflectance spectra of Ir$_{0.95}$Pt$_{0.05}$Te$_2$ single crystals obtained on samples of $3\times$ 2.8 mm$^2$ in $ab$-plane dimensions. (b) Optical conductivity $\sigma_1(\omega)$ spectra at various temperatures up to 18000 \cm.}
\end{figure}

\begin{figure}
\includegraphics[clip,width=3.7in]{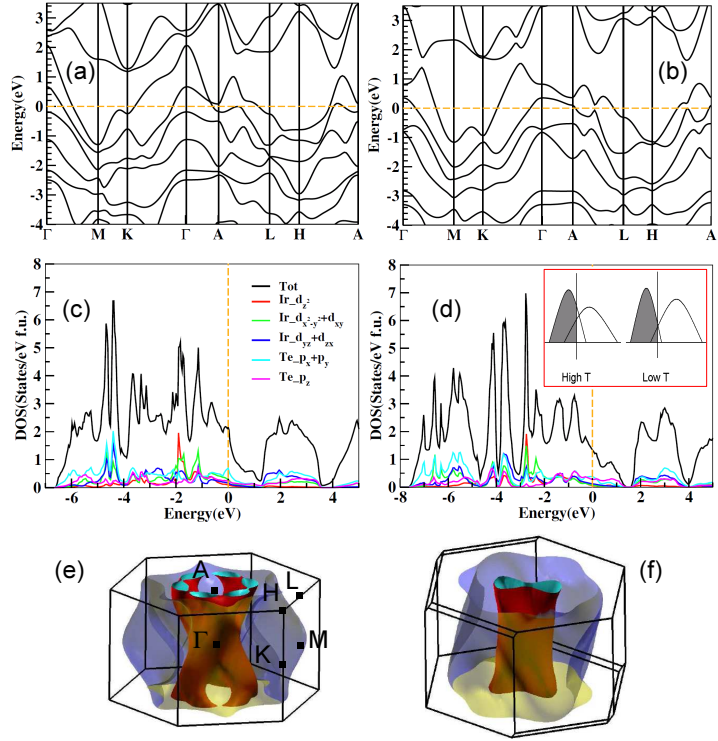}
\caption{(Color online) The calculated band structures ((a) and (b)), the density of states (DOS) including partial DOS contributions from different Ir 5d and Te 5p orbitals ((c) and (d)), and resultant Fermi surfaces ((e) and (f)) in terms of the reported structural data \cite{resistance} at high and low temperature phases, respectively. }
\end{figure}

\end{document}